\documentclass[conference,10pt,a4paper]{IEEEtran}
\usepackage[cmex10]{amsmath}
\usepackage{amsfonts,bm}
\usepackage{algorithmic,array,cite,mdwmath,mdwtab,eqparbox}
\ifCLASSINFOpdf
  \usepackage[pdftex]{graphicx}
  \graphicspath{{figures/pdf/}{figures/png/}{figures/jpg/}}
  \DeclareGraphicsExtensions{.pdf,.jng,.jpeg}
\else
  \usepackage[dvips]{graphicx}
  \graphicspath{{figures/eps2/}}
  \DeclareGraphicsExtensions{.eps}
\fi
\usepackage[caption=false,font=footnotesize]{subfig}
\usepackage{fixltx2e}
\usepackage{url}
\usepackage{tikz}
\newtheorem{theorem}{Theorem}

\newcounter{exNo}
\def\vec#1{{\bf#1}}

\def\ket#1{| #1 \rangle}
\def\bra#1{\langle #1 |}

\def\ave#1{\langle #1 \rangle}

\def\norm#1{\| #1 \|}

\def\Tr{\operatorname{Tr}}

\def\diag{\operatorname{diag}}

\def\H{\mathcal{H}}

\def\ONE{\mathbb{I}}

\def\DFT{\mbox{\rm DFT}}

\begin{document}
\title{Quantum System Identification:
Hamiltonian Estimation using Spectral and Bayesian Analysis}

\author{\IEEEauthorblockN{Sophie G Schirmer\IEEEauthorrefmark{1},
Frank C Langbein\IEEEauthorrefmark{2} 
}
\IEEEauthorblockA{\IEEEauthorrefmark{1} 
     Department of Applied Maths \& Theoretical Physics, Univ. of Cambridge,\\ 
     Wilberforce Rd, Cambridge, CB3 0WA, UK;  
     Email: sgs29@cam.ac.uk}
\IEEEauthorblockA{\IEEEauthorrefmark{2} 
     School of Computer Science, Cardiff University,\\
     5 The Parade, Cardiff, CF24 3AA, UK;
     Email: F.C.Langbein@cs.cardiff.ac.uk}
}
\maketitle

\begin{abstract}
Identifying the Hamiltonian of a quantum system from experimental data
is considered. General limits on the identifiability of model
parameters with limited experimental resources are investigated, and a
specific Bayesian estimation procedure is proposed and evaluated for a
model system where a-priori information about the Hamiltonian's
structure is available.
\end{abstract}

\section{Introduction}

At a fundamental level nature is governed by the laws of quantum
mechanics, but until recently such phenomena were mostly a curiosity
studied by physicists. However, significant advances in theory and
technology are increasingly pushing quantum phenomena into the realm
of engineering, as building blocks for novel technologies and
applications from chemistry to computing. E.g., advances in laser
technology enable ever more sophisticated coherent control of atoms,
molecules and other quantum systems. Recent advances in
nanofabrication have made it possible to create nanostructures such as
quantum dots and quantum wells that behave like artificial atoms or
molecules and exhibit complex quantum behaviour. Cold-atom systems and
the creation of Bose condensates demonstrate that even macroscopic
systems can exhibit quantum coherence.

Harnessing the potential of quantum systems is a challenging task,
requiring exquisite control of quantum effects and system designs that
are robust with regard to fabrication imperfection, environmental
noise and loss of coherence. Although significant progress has been
made in designing effective controls, most control design is
model-based, and available models for many systems do not fully
capture their complexity. Model parameters are often at best
approximately known and may vary, in particular for engineered systems
subject to fabrication tolerances. Experimental system identification
is therefore crucial for the success of quantum engineering. While
there has been significant progress in quantum state identification
and quantum process tomography, we require dynamic models if we wish
to control a system's evolution. Furthermore, effective protocols must
take into account limitations on measurement and control resources for
initial device characterization. This presents many challenges, from
determing how much information can be obtained in a given setting to
effective and efficient protocols to extract this information. Here we
illustrate some problems and solutions for the case of identifying the
dynamics of a three-level system.

\section{Identifiability of Model Parameters}

One of the first questions to consider before attempting to find
explicit protocols for experimental system identification is clearly
what information we can hope to extract about a given system with a
certain limited set of resources. For instance, given a system with a
Hilbert space of dimension $N$, it is well known that the ability to
prepare and measure the system in a set of computational basis states
$\{\ket{n}: n=1\ldots,N\}$ is insufficient for quantum process
tomography, even if the process is
unitary~\cite{JMO44p2455,PRL78p390}. However, recent work shows that a
substantial amount of information about the generators of the dynamics
can be obtained for Hamiltonian
\cite{PRA69n050306(R),qph0409107,PRA71n062312,PRA80n022333} and even
dissipative systems~\cite{PRA73n062333,NJP9p384,BIRS2007,qph0911_1367}
at least generically, by mapping the evolution of the computational
basis states stroboscopically over time. More precisely, this is done
by determining the probabilities that a measurement of the observable
$M=\diag(m_1,\ldots,m_N)$ produces the outcome $m_\ell$ after the
system was initialized in the computational basis state $\ket{k}$ and
allowed to evolve for time $t$ for a number of different times $t_n$.
This begs the question how much information we can hope to obtain in
general from such experiments. In this paper we consider Hamiltonian
systems, whose evolution is governed by the Schrodinger equation
$i\hbar\dot{U}(t,t_0)=HU(t,t_0)$ with a fixed Hamiltonian $H$ and
$U(t_0,t_0)=\ONE$, for which we have
$p_{k\ell}(t)=|\bra{\ell}U(t,t_0)\ket{k}|^2$.

\begin{theorem}
Let $H$ and $M$ be Hermitian operators representing the Hamiltonian and
the measurement, respectively, and let $\rho_0$ be a positive operator
with $\Tr(\rho_0)=1$ representing the initial state of the system.  If
$M$, $H$ and $\rho_0$ are simultaneously blockdiagonalizable, i.e., 
there exists a decomposition of the Hilbert space $\H=\oplus_{s=1}^{S>1} 
\H_s$ such that
\begin{equation}
  M=\diag(M_s), \quad H=\diag(H_s), \quad \rho_0=\diag(\rho_s), 
\end{equation}
where $M_s$, $H_s$ and $\rho_s$ are operators on the Hilbert spaces
$\H_s$, then we can at most identify $H$ up to $\sum_s \lambda_s\ONE_s$,
where $\ONE_s$ is the identity on the subspace $\H_s$.
\end{theorem}

\begin{IEEEproof}
If $H$ is block-diagonal then any initial state $\rho_0$ starting in a
subspace $\H_s$ must remain in this subspace. Thus, the dynamics on each
subspace is independent, $U(t)=\otimes_s U_s(t)$ with $U_s(t)=
e^{-it H_s}$.  Per hypothesis $M$ and $\rho_0$ are also blockdiagonal, 
so $\Tr[M U(t)\rho_0 U(t)^\dag] = \sum_s \Tr[M_s U_s(t)
\rho_s U_s(t)^\dag]$.  If $\tilde{H}=H+\sum_s \lambda_s \ONE_s$ then 
$\tilde{U}(t)=\otimes_s \tilde{U}_s(t)$ with 
$\tilde{U}_s(t)=e^{-it\lambda_s} U_s(t)$. Thus,
$\Tr[M \tilde{U}(t) \rho_0 \tilde{U}(t)^\dag]$
$=\sum_s \Tr[M_s e^{-it\lambda_s}U_s(t) \rho_s e^{it\lambda_s}U_s(t)^\dag]$ 
$=\sum_s \Tr[M_s U_s(t) \rho_s U_s(t)^\dag]$
shows that $H$ and $\tilde{H}$ are indistinguishable.
\end{IEEEproof}

Thus, there are some limitations on the maximum amount of information we
can obtain about the system by initializing and measuring the system in
a fixed computational basis.  In particular, if $H$ and $M$ commute, we
can infer that $H$ and $M$ are simultaneously diagonalizable, and
assuming the eigenvalues $m_\ell$ of $M$ are distinct, this fixes the
Hamiltonian basis, i.e., we have $H=\sum_\ell \lambda_\ell \Pi_\ell$,
where $\Pi_\ell$ is the projector on the eigenspace of $M$ corresponding
to $m_\ell$, i.e., the computational basis state $\ket{\ell}$.  However,
no information about the eigenvalues $\lambda_m$ or the transition
frequencies $\omega_{k\ell}=\lambda_\ell-\lambda_k$ can be obtained by
measuring $p_{k\ell}(t)$, all of which are constant in this case.

Maximum information about the Hamiltonian can be obtained if $H$ and $M$
are not simulataneously block-diagonalizable.  This is the generic case, and
in this case we can identify $H$ at most up to a diagonal unitary matrix
$D=(1,e^{i\phi_2},\ldots,e^{i\phi_N})$ and a global energy shift
$\lambda_0\ONE$, i.e., $\tilde{H} \simeq H = D^\dag \tilde{H} D +
\lambda_0\ONE$, as was noted in~\cite{PRA80n022333}.  The term
$\lambda_0\ONE$ is generally physically insignificant as it gives rise
only a global phase factor $\tilde{U}(t,0)=e^{-i t
(H+\lambda_0\ONE)}=e^{-i\lambda_0 t} e^{-i tH} e^{-i\lambda_0 t}U(t,0)$,
which is generally unobservable, as the abelian phase factors cancel,
$\rho(t)=U(t,0)\rho_0 U(t,0)^\dag=\tilde{U}(t,0)\rho_0
\tilde{U}(t,0)^\dag$ for any $\rho_0$.  The diagonal unitary matrix $D$
represents the freedom to redefine the measurement basis states,
$\ket{n}\mapsto e^{i\phi_n} \ket{n}$ as
$\Pi_n=\ket{n}\bra{n}=e^{i\phi_n}\ket{n}\bra{n}e^{-i\phi_n}$.  The
phases $\phi_n$ cannot be ignored in general but in certain special
cases they can be effectively eliminated.  For example, if $H$ is known
to be real-symmetric, a common case in physics, then we can choose all
basis vectors to be real and restrict $e^{i\phi_n}$ to $\pm 1$.
Moreover, if the off-diagonal elements in the computational basis are
known to be real and positive, $H_{k\ell}=\bra{k}H\ket{\ell}= |\bra{k}
H\ket{\ell}|$, then $|\bra{k}H\ket{\ell}|=|\bra{k}\tilde{H}\ket{\ell}|$
with $\tilde{H}$ as above.  Hence, with this additional constraint
the Hamiltonian is effectively uniquely determined (up to a global
energy level shift and global inversion of the energy levels).

A constructive procedure for reconstructing a generic unknown
Hamiltonian from stroboscopic measurements of the observables
$p_{k\ell}(t)$ at fixed times $t=t_n$ using Bayesian parameter
estimation techniques was also given in~\cite{PRA80n022333}.

\section{Identification with a-priori information}

The previous section shows that when essentially no a-priori information
about the Hamiltonian is available then even measurement of all the
observables $p_{k\ell}(t)$ is not sufficient to uniquely determine the
Hamiltonian.  However, in many cases some a-priori knowledge about the
system is available.  For instance, the transition frequencies
$\omega_{\mu\nu} =\lambda_\nu-\lambda_\mu$ of the system, where
$\lambda_\nu$ are the eigenvalues of the Hamiltonian $H$, may be known
from available spectroscopic data, and we may be able to infer basics
such as the level structure and allowed transitions from fundamental
physical principles. In such cases the identification problem can be
substantially simplified and far less information may be required.

As a specific simple example, consider a three-level system with known
transition frequencies $\omega_{12}$ and $\omega_{23}$ and no direct
transitions between states $\ket{1}$ and $\ket{3}$ subject to external
fields driving the $(1,2)$ and $(2,3)$ transitions, respectively. If
our computational/measurement basis coincides with the eigenbasis of the
undriven system, then we know that the Hamiltonian of the driven system
must be of the form $H=H_0+f(t)H_1$ with $H_0=\diag(0,\omega_{12},
\omega_{12}+\omega_{23})$ and 
$H_1 = \left[\begin{smallmatrix} 
            0 & d_1 & 0 \\ 
            d_1 & 0 & d_2 \\ 
            0 & d_2 & 0 
	    \end{smallmatrix} \right]$, 
i.e., we have only two unknowns, $d_1$ and $d_2$.  If we take the field
to be of the form $f(t)=A_1 \cos(\omega_{12} t)+A_2\cos(\omega_{23}t)$,
i.e., consisting of two components that resonantly drive the $(1,2)$ and
$(2,3)$ transition, then transforming to a rotating frame and making the
rotating wave approximation (RWA), we obtain an effective Hamiltonian
$H_{\rm eff} = \left[\begin{smallmatrix} 
		      0 & \Omega_1 & 0 \\ 
		      \Omega_1 & 0 & \Omega_2 \\ 
		      0 & \Omega_2 & 0
\end{smallmatrix} \right]$, 
where $\Omega_k=d_k A_k/2\hbar$ for $k=1,2$.  If the field amplitudes
$A_k$ are constant, this Hamiltonian is constant and we could use the
general protocol in \cite{PRA80n022333} to fully characterize the
dynamics by stroboscopically measuring the probabilities $p_{k\ell}(t)$
for $k,\ell=1,2,3$ at sufficiently many times $t_n$.  This requires the
ability to initialize the system in all three basis states $\ket{k}$ and
measure the populations of all three states.  Due to conservation of
probability $\sum_\ell p_{k\ell}=1$ and symmetry $p_{k\ell}=p_{\ell k}$,
the requirements can be reduced to initialization and measurement in two
basis states, e.g., $\ket{1}$ and $\ket{3}$, as the remaining
probabilities can be inferred from the other two, but we can do even
better by using all the information available.

We shall assume $\Omega_1$ and $\Omega_2$ are real and positive.  For
notational convenience, let $\Omega=\sqrt{\Omega_1^2+\Omega_2^2}$ and
$\alpha=\arctan(\Omega_2/\Omega_1)$ be the polar coordinates of the
vector $(\Omega_1,\Omega_2)$, i.e., $\Omega_1=\Omega\cos\alpha$ and
$\Omega_2=\Omega\sin\alpha$ with $\Omega\in\mathbb{R}_0^+$ and
$\alpha\in[0,\pi/2]$.  Then $U(t,0)=\exp(-it H_{\rm eff})$ is
\begin{equation}
  \begin{bmatrix}
  c^2\cos(\Omega t)+s^2 &-ic\sin(\Omega t) & cs[\cos(\Omega t)-1]\\
  -i c\sin(\Omega t) & \cos(\Omega t) & -is\sin(\Omega t)\\
  cs[\cos(\Omega t)-1] & -is\sin(\Omega t)& s^2\cos(\Omega t)+c^2
 \end{bmatrix}
\end{equation}
where $c=\cos\alpha$ and $s=\sin\alpha$.  This shows immediately that
a single measurement trace $p_{k\ell}(t)=|\bra{\ell}U(t,0)\ket{k}|^2$ 
except $p_{22}(t)$ contains information about both parameters and thus 
should be sufficient to fully identify the Hamiltonian.  Specifically, 
if we choose to measure $p_{11}(t)$ we obtain
\begin{align*}
  p_{11}(t)&= \textstyle x^2\cos^2(\Omega t)+ 2x(1-x)\cos(\Omega t)+(1-x)^2\\
           &= \textstyle \frac{x^2}{2} \cos(2\Omega t) 
              +2 x(1-x)\cos(\Omega t) + (1-x)^2+\frac{x^2}{2},
\end{align*}
using $\cos^2(\Omega t)=\frac{1}{2}[\cos(2\Omega t)+1]$ and setting
$x=c^2=1-s^2$.  This shows that there are three frequency components
$0$, $\Omega$ and $2\Omega$, whose amplitudes determine $\alpha$.

\section{Efficient Parameter Estimation}

\begin{figure}
\includegraphics[width=\columnwidth]{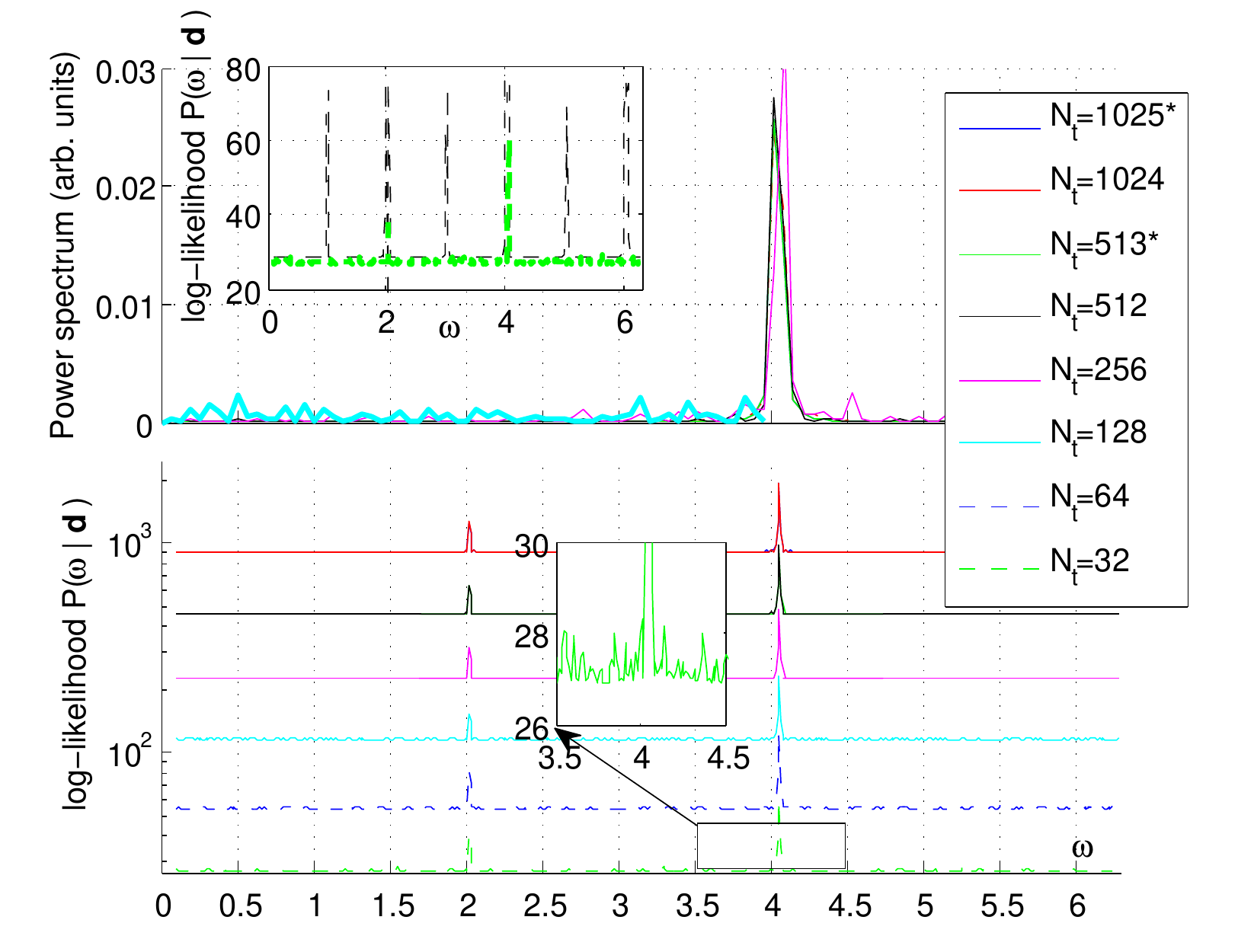} \caption{Power spectra and
log-likelihood for a test system with $\Omega=4.0484$ for data sampled
at different times $\vec{t}$ in $[0,100]$.  For $N_t\ge 128$ the power
spectra have a single peak in the plotted range, which is a reasonable
estimate for $\Omega$.  For $N_t=64$ and below, the main peak is outside
the range of the power spectrum and the former no longer contains any
useful information.  Yet, the log-likelihood still has a clearly
identifiable global maximum at $\Omega$ even for data vectors with as
few as 32 data points, provided a non-uniform sampling is used.  For
uniform sampling with $N_t=32$ the top inset shows that
$P(\omega|\vec{d})$ has many peaks of approximately equal height due to
aliasing effects (dashed black line).}  \label{fig1}
\end{figure}

The form of $p_{11}(t)$ suggests Fourier analysis to determine the
parameters $\Omega$ and $\alpha$, e.g., by identifying the non-zero
Fourier components.  The highest frequency peak will be at $2\Omega$ and
the corresponding peak amplitude $a_2=x^2/2$ uniquely determines
$x=\sqrt{2a_2}$.  In some cases (as in the example shown in
Fig.~\ref{fig1}) there may be only one clearly identifiable non-zero
peak in the power spectrum, which could correspond to either $\Omega$ or
$2\Omega$.  This problem can in principle be overcome by estimating $x$
from the average signal $\ave{p_{11}(t)}=a_0(x)=1-2x+\frac{3}{2}x$, from
which we can obtain estimates for the coefficients
$a_2(x)=\frac{1}{2}x^2$ and $a_1(x)=2x(1-x)$.  If $a_2\gg a_1$ then we
identify the non-zero-frequency peak with $2\Omega$, otherwise with
$\Omega$.

Alternatively, we can estimate the base frequency $\Omega$ and the
signal amplitudes using a Bayesian approach.  The signal in our case is
a linear combination of the basis functions $g_0=1$, $g_1(t)=\cos(\Omega
t)$ and $g_2(t)=\cos(2\Omega t)$.  Following standard techniques, we
maximize the log-likelihood function~\cite{PRA80n022333,88Bretthorst}
\begin{equation}
  P(\omega|\vec{d}) \propto \frac{m_b-N_t}{2} \log_{10}
     \left[1-\frac{m_b \ave{\vec{h}^2}} {N_t\ave{\vec{d}^2}}\right],
\end{equation}
where $m_b$ is the number of basis functions, $m_b=3$ in our case, $N_t$
is the number of data points, and
\begin{equation}
 \ave{\vec{d}^2}=\frac{1}{N_t} \sum_{n=0}^{N_t-1} d_n^2, \quad
 \ave{\vec{h}^2}=\frac{1}{m_b} \sum_{m=0}^{m_b-1} h_m^2, 
\end{equation}
where the elements $h_m$ of $(m_b,1)$-vector $\vec{h}$ are projections
of the $(1,N_t)$-data vector $\vec{d}$ onto a set of orthonormal basis
vectors derived from the non-orthogonal basis functions $g_m(t)$
evaluated at the respective sample times $t_n$.  Concretely, setting
$G_{mn}=g_m(t_n)$, let $\lambda_m$ and $\vec{e}_m$ be the eigenvalues
and corresponding (normalized) eigenvectors of the $m_b\times m_b$
matrix $G G^\dag$ with $G=(G_{mn})$, and let $E=(e_{m'm})$ be a matrix
whose columns are $\vec{e}_m$.  Then we have $H=V G$ and $\vec{h}=H
\vec{d}^\dag$ with $V =\diag(\alpha_m^{-1/2}) E^\dag$, and the
corresponding coefficient vector is $\vec{a}=\vec{h}^\dag V$.

In our case the $P(\omega|\vec{d})$ is a function of a single frequency
$\omega$ and $\Omega$ is the frequency for which $P(\omega|\vec{d})$
achieves its global maximum.  If $\vec{a}(\Omega)$ is the corresponding
coefficient vector, we can obtain the best estimate for $x=\cos^2\alpha$
and thus $\alpha$ by minimizing $\norm{\vec{a}(x)-\vec{a}(\Omega)}$ with
$a_m(x)$ as defined above.  Thus, the problem of finding the most likely
model $(\Omega,\alpha)$ is reduced to finding the global maximum of
$P(\omega|\vec{d})$.  Unfortunately, this is not an easy task as
$P(\omega|\vec{d})$ is sharply peaked and can have many local extrema
and a substantial noise floor depending on the number and accuracy of
the data points.  One way to circumvent this problem is to use the peaks
in the discrete Fourier spectrum $\DFT(\vec{d})$ of the data $\vec{d}$
as input for a gradient-based optimization of $P(\omega|\vec{d})$.  To
make the peak detection simpler and more robust, especially when the
data is noisy, we find the position $\omega_0$ of the highest peak in
the rescaled power spectrum $F(\omega)= 20 \log_{10}
[|\DFT[\vec{d}-\ave{\vec{d}}]|^2+1]$, which should correspond to either
$\Omega$ or $2\Omega$, and then find the location of the maxima
$\omega_1$ and $\omega_2$ of $P(\omega|\vec{d})$ in the intervals
$I_1=[\omega_0-\Delta\omega,\omega_0+\Delta\omega]$ and
$I_2=[\frac{1}{2}\omega_0-\Delta\omega,\frac{1}{2}\omega_0+\Delta\omega]$,
where $\Delta\omega$ depends on the resolution of the discrete Fourier
transform, e.g., $\Delta\omega\approx 2\pi/T$ for regularly sampled
data.  We take the best estimate $\omega_3$ for the system frequency 
$\Omega$ to be $\omega_1$ if $P_1>P_2$, and $\omega_2$ otherwise, where
$P_j=P(\omega_j|\vec{d})$ for $j=1,2$.  If $P_1$ and $P_2$ differ by 
less than a certain amount we can flag the system suggesting
that more data is needed for reliable discrimination.

To test this strategy 30 Hamiltonians $H(\Omega_k,\alpha_k)$ with
$\Omega_k\in [0,2\pi]$ and $\alpha_k\in[0,\frac{\pi}{2}]$ and a range
of sampling time vectors $\vec{t}=(t_n)$ with $t_n \in [0,100]$ were
generated with the number of samples $N_t$ ranging from $2^{10}$ to
$2^5$. Regular and irregular time vector samplings were considered,
where for irregular samples a (fast) non-uniform Fourier transform was
used~\cite{Greengard}. For each test system and time vector
$\vec{t}_\ell$, noisy data vectors $\vec{d}$ were generated by
simulating actual experiments, noting that in a laboratory experiment
each data point $d_n$ would normally be estimated by initializing the
system in state $\ket{1}$, letting it evolve for time $t_n$, and
performing a projective measurement $P_1=\ket{1}\bra{1}$, whose
outcome is random, either $0$ or $1$. To estimate the probability
$p_{11}(t_n)$ the experiment is repeated many times and $p_{11}(t_n)$
approximated by the relative frequency $d_n$ of $1$'s. The simplest
approach is to use a fixed number of experiment repetitions $N_e$ for
each time $t_n$, but noting that the uncertainty of the estimate $d_n$
of $p_{11}(t_n)$ is $N_e^{-1/2}$ shows that it is advantageous to
adjust the number of repetitions $N_e$ for each time $t_n$ to achieve
a more uniform signal-to-noise ratio. Specifically, for each data
point we sample until $d_n \sqrt{N_e}\approx 10$ or we reach a maximum
number of repetitions (here $10^4$). Although the projection noise for
a single data point is Poissonian, the overall error distribution for
a large number of samples is roughly Gaussian, justifying the use of a
Gaussian error model in the Bayesian analysis.

As the resolution of the discrete Fourier transform and hence the
scaled power spectrum is approximately $2\pi/T$, and generally
somewhat less for irregular sampling, the uncertainty in the peak
positions of the power spectrum will generally be at least $\pi/T$,
limiting the accuracy of the frequency estimates, in our case to
$\approx 0.0314$, regardless of the number of data points. This is
evident in Fig.~\ref{fig1}, which shows that the peak in power
spectrum is relatively broad, compared to the peak in the likelihood
function. Furthermore, the frequency range covered by the power
spectrum depends on the sampling frequency, or the number of data
points $N_t$, with the largest discernible frequency approximately
$N_t\pi/T$. If the system frequency $\Omega$ is outside this range
covered by the power spectrum, we are unable to detect it. For
example, for a system with $\Omega=4.0484$, we require
$N_t\pi/T>\Omega$ and thus $N_t>128$ data points (see
Fig.~\ref{fig1}). If $T$ and $N_t$ are sufficiently large to avoid
such problems, the location $\omega_0$ of the global maximum of the
power spectrum usually provides a good starting point for finding the
global optimum of the log-likelihood function but we can generally
substantially improve the frequency estimates using the likelihood. Of
14440 data sets analyzed (30 test systems sampled at different times)
$\omega_0$ differed by less than 1\% from the true system frequency
$\Omega$, or $2\Omega$, i.e., $E(\omega_0)<0.01$ with
$E(\omega_0)=\min\{|\omega_0-\Omega|/\Omega,|\omega_0-2\Omega|/2\Omega\}$
in about half (7321) the cases. For almost all failed cases the number
of data points was too small and $\Omega$ outside the range of the
power spectrum. Even when restricted to the successful cases as
defined above, the median of $E(\omega_0)$ was $0.0035$, while the
median of the relative error $E_1(\omega_3)=|\omega_3-\Omega|/\Omega$
of the final estimate $\omega_3$ obtained by maximizing the likelihood
was $6.9\times 10^{-6}$.

We also considered finding the global maximum of the likelihood by
other means, especially in those cases for which the power spectrum
does not provide a useful initial frequency estimator. Since we have a
function of a single parameter and evaluation of the likelihood,
especially when the number of data points is small, is not expensive,
it is possible to find the global maximum simply by exhaustive search.
Interestingly, we found that log-likelihood still had a clearly
identifiable global maximum in many cases even when the number of data
points $N_t$ was far below the minimum number of sample points
required to detect a peak in the power spectrum. E.g., for the system
shown in Fig.~\ref{fig1}, the likelihood function still has a sharp
peak around the system frequency $\Omega$ even if the number of
samples is reduced to $32$, while the peak is no longer detectable in
the power spectrum even for $N_t=128$ samples. However, as we reduce
the number of samples additional peaks in the likelihood function tend
to emerge at multiples or fractions of $\Omega$, as shown in the top
inset of Fig.~\ref{fig1}. This means that we can no longer
unambiguously identify the true frequency $\Omega$. Such aliasing
problems leading to sampling artefacts in the data analysis can be
sustantially reduced by avoiding uniform sampling at equally spaced
times (cf Fig.~\ref{fig1}, top inset). In particular low-discrepancy
sequences have been introduced with the aim to create a sampling with
minimal regular patterns causing sampling artefacts, but also
minimising the average gap between the samples for a fixed number of
samples~\cite{Niederreiter}. Here in particular we use a stratified
sampling strategy, where a point is placed in each stratum of a
regular grid according to a uniform probability distribution. This
\emph{may} be improved further using other low-discrepancy
sequences~\cite{Quinn}. The results are relevant as a significant
reduction in the number of data points required reduces experimental
overheads substantially. This comes at additional computational costs,
as finding the global maximum of the likelihood function for irregular
samplings with very few data points forms a hard optimization problem.
Several standard optimization algorithms (simple pattern search and
stochastic gradient decent) failed to reliably detect the global
optimum, and exhaustive search had to be used.

\section{Concluding discussion}

We have considered Hamiltonian identification using stroboscopic
measurement data of a fixed observable. If the system can only be
initialized in the measurement basis states then a completely unknown
Hamiltonian cannot be uniquely identified even if we can measure the
population of all basis states as a function of time. If a-priori
information is available, however, complete identification of the
system parameters is often possible with substantially reduced
resources. We have illustrated this for the case of a three-level
system where we can only monitor the population of state $\ket{1}$
over time, starting in $\ket{1}$, without the possibility of dynamic
control or feedback as was considered in \cite{arXiv0903_1011}. The
results may be applicable to improve the efficiency of identification
schemes for other systems. E.g., recent work on system identification
for spin networks \cite{PRA79n020305(R),NJP11n103019} has shown that
the relevant Hamiltonian parameters of a spin chain can be identified
by mapping the evolution of the first spin and Fourier analysis, but
the scheme requires repeated quantum state tomography of the first
spin for many times $t_n$, which is experimentally expensive.

\section{Acknowledgments}

SGS acknowledges funding from EPSRC ARF Grant EP/D07192X/1, the EPSRC
QIP Interdisciplinary Research Collaboration (IRC), Hitachi and NSF
Grant PHY05-51164.  FCL acknowledges funding for RIVIC One Wales
national research centre from WAG.


\begin{thebibliography}{99}

\bibitem{JMO44p2455}
I. L. Chuang and M. A. Nielsen, J. Mod. Opt. \textbf{44}, 2455-2467 (1997).

\bibitem{PRL78p390}
J. F. Poyatos, J. I. Cirac and P. Zoller, Phys. Rev. Lett. \textbf{78}, 390 (1997).

\bibitem{PRA69n050306(R)}
S.~G.~Schirmer, A.~Kolli, D.~K.L.~Oi, Phys.~Rev.~A \textbf{69}, 050306(R) (2004).

\bibitem{qph0409107} 
S. G. Schirmer, A. Kolli, D. K. L. Oi, J. H. Cole, 
In: Proc. 7th Int. Conf. QCMC, Glasgow 25-29 July 2004 (AIP 2004).

\bibitem{PRA71n062312}
J. H. Cole \emph{et al.} 
Phys. Rev. A \textbf{71}, 062312 (2005).

\bibitem{PRA80n022333}
S. G. Schirmer and D. K. L. Oi, Phys. Rev. A \textbf{80}, 022333 (2009)

\bibitem{PRA73n062333}
J. H. Cole \emph{et al.}
Phys. Rev. A \textbf{73}, 062333 (2006).

\bibitem{NJP9p384}
S. J. Devitt \emph{et al.}
New J. Phys. \textbf{9}, 384 (2007).

\bibitem{BIRS2007}
S. G. Schirmer, D. K. L. Oi and S. J. Devitt, 
J. Phys.: Conf. Series \textbf{107} 012011 (2008)

\bibitem{qph0911_1367}
S. G. Schirmer and D. K. L. Oi, arXiv:0911.1367 (2009)

\bibitem{88Bretthorst}
G. Larry Bretthorst, Bayesian Spectrum Analysis and Parameter Estimation
(Springer, Berlin, 1998)

\bibitem{arXiv0903_1011}
Z. Leghtas, M. Mirrahimi, P. Rouchon, arXiv:0903.1011 (2009)

\bibitem{PRA79n020305(R)}
D. Burgarth, K. Maruyama, F. Nori,
Phys. Rev. A \textbf{79}, 020305(R) (2009)

\bibitem{NJP11n103019}
Daniel Burgarth, Koji Maruyama, New J. Phys. \textbf{11}, 103019 (2009)

\bibitem{Niederreiter} 
H. Niederreiter, Random Number Generation and Quasi-Monte Carlo Methods
(SIAM Review, 1992)

\bibitem{Quinn}
J. A. Quinn, F. C. Langbein, R. R. Martin, G. Elber,
Springer LNCS \textbf{4077}, 465-484 (2006).


\bibitem{Greengard}
L. Greengard, J. Lee, SIAM Review \textbf{46}(3), 443-454 (1993).


\end{thebibliography}
\end{document}